\documentclass{aa}

\usepackage{color}
\usepackage[varg]{txfonts}
\usepackage{graphics}
\usepackage{hyperref}

\newcommand{\be}{\begin{equation}}
\newcommand{\ee}{\end{equation}}
\newcommand{\beq}{\begin{eqnarray}}
\newcommand{\eeq}{\end{eqnarray}}
\newcommand{\bc}{\begin{center}}
\newcommand{\ec}{\end{center}}
\newcommand{\unit}[1]{\mbox{\boldmath $\hat{#1}$}}
\newcommand{\bm}[1]{\mbox{\boldmath ${#1}$}}

\newcommand{\req}{R_{\mathrm{e}}}
\newcommand{\rpol}{R_{\mathrm{p}}}
\newcommand{\ftot}{F^{\mathrm{tot}}}

\newcommand{\msun}{{M}_{\sun}}

\newcommand{\diff}{{\rm d}}

\begin{document}

\title{Oblate Schwarzschild approximation for polarized radiation from rapidly rotating  neutron stars}

\titlerunning{Polarized oblate Schwarzschild approximation}

\author{Vladislav~Loktev\inst{1,2}
\and Tuomo~Salmi\inst{1}
\and Joonas~N\"attil\"a\inst{4,5}
\and Juri~Poutanen\inst{1,2,3}}

\institute{Department of Physics and Astronomy, FI-20014 University of Turku, Finland \\ \email{vladislav.loktev@utu.fi, juri.poutanen@utu.fi}
\and Space Research Institute of the Russian Academy of Sciences, Profsoyuznaya str. 84/32, 117997 Moscow, Russia 
\and Nordita, KTH Royal Institute of Technology and Stockholm University, Roslagstullsbacken 23, SE-10691 Stockholm, Sweden
\and Physics Department and Columbia Astrophysics Laboratory, Columbia University, 538 West 120th Street New York, NY 10027
\and Center for Computational Astrophysics, Flatiron Institute, 162 Fifth Avenue, New York, NY 10010, USA
}

\date{Received 10 August 2020 / Accepted 08 September 2020	}

\abstract{We have developed  a complete theory for the  calculation of the observed Stokes parameters for radiation emitted from the surface of a rapidly rotating neutron star (NS) using the oblate Schwarzschild approximation. 
We accounted for the rotation of the polarization plane due to relativistic effects along the path from the stellar surface to the observer.
The results were shown to agree with those obtained by performing full numerical general relativistic ray-tracing with the \textsc{arcmancer} code. 
We showed that the obtained polarization angle (PA) profiles may differ substantially from those derived for a spherical star.
We demonstrated that assuming incorrect shape for the star can lead to biased constraints for NS parameters when fitting the polarization data.
Using a simplified model, we also made a rough estimate of how accurately the geometrical parameters of an accreting NS can be determined using the X-ray polarization measurements of upcoming polarimeters like   the  \text{Imaging X-ray Polarimeter Explorer} (\text{IXPE}) or \text{the enhanced X-ray Timing and Polarimetry} (\text{eXTP}) mission.
}


\maketitle

\section{Introduction}\label{sec:intro}

Millisecond pulsars (MSPs) are rapidly rotating neutron stars (NSs) with relatively low magnetic field strengths. 
Many of the known MSPs are found in binary systems where the in-falling gas from the companion star spins up the central NS to extreme angular velocities \citep{RS82,ACR82}.
During the accretion phase, matter falls on the NS surface near the magnetic poles creating hotspots that emit polarized X-ray radiation. 
X-ray polarization from accretion-powered millisecond pulsars (AMPs) will be soon measured by  upcoming space observatories like the \text{Imaging X-ray Polarimeter Explorer} (\text{IXPE}) \citep{Weisskopf16} and the \text{enhanced X-ray Timing and Polarimetry} mission (\text{eXTP}) \citep{Zhang19,Watts19}. 
The pulse profiles contain information about parameters of the NS, such as its mass and radius  \citep[e.g.,][]{PG03,Watts16,SNP18,Bogdanov19L26}. 
The relation between these parameters can give constraints on the equation of state of extremely dense matter constituting the inner parts of NSs \citep[see, e.g.,][]{lindblom1992,Lattimer12ARNPS}.

The core theory on pulse profile modeling was presented in \cite{PFC83}, and was later extended for rapidly rotating spherical NSs by \citet{PG03} and \citet{PB06}.
They used the  Schwarzschild+Doppler approximation (S+D), where the effects of light bending and gravitational redshift are computed in Schwarzschild metric and special relativistic effects (aberration, Doppler effect) are included separately. 
In addition to the modulation of the X-ray flux, \citet{VP04} modeled the polarization degree (PD) and polarization position angle (PA) from an infinitely small hotspot (or two antipodal spots) on the surface of rapidly rotating spherical NS.
\citet{poutanen20b} presented a derivation of the formulae from \citet{VP04} using a  common vector formalism and also studied the effects of rapid rotation on the polarization profile.

The effects of the deviation of the metric and stellar shape from  spherical  were considered by \citet{BRR00},  \citet{CMLC07},  \citet{Baubock.etal:12}, and recently by \citet{VBR18}, \citet{NP18}, and \citet{PMN18}. 
For realistic NS equations of state and spins, the dominant effect comes from the oblateness of the NS surface. 
This prompted \citet{CMLC07} and \citet{MLC07} to suggest an oblate Schwarzschild (OS) approximation to describe the flux modulation from a rapidly rotating NS. 
This approximation accounts for the deviation of the NS shape from a sphere, but treats the surrounding space-time using the spherically symmetric Schwarzschild metric.
The corrected OS formalism, also valid   for  finite spot size, was recently presented by \citet{SNP18}, \citet{Bogdanov19L26}, and \citet{SP20}, but without considering polarization. 
Here we combine the techniques described in those papers to compute the pulse profiles with the methods suggested by \citet{VP04} and \citet{poutanen20b} to compute polarization.
With this combination we develop the complete theory needed to calculate  the Stokes vector light curves from rapidly rotating oblate NSs in the OS approximation. 
We derive analytical expressions for apparent PA for a point source at the oblate surface of a rapidly rotating NS (which are also applied to calculate the PA from a finite-sized spot).
Then we use these formulae to model pulse profiles and polarization and study the importance of the oblateness effects in such modeling.

The remainder of this paper is structured as follows. 
In Sect.~\ref{sec:theory}, we present the theory that describes the transfer of polarized radiation from one or two hotspots at the surface of an oblate NS to the observer at infinity. 
In Sect.~\ref{sec:results} we first confirm the accuracy of our model by comparing the polarization pulse profiles to those calculated with the general relativistic polarized radiative transfer code \textsc{Arcmancer} \citep{PMN18} using oblate star shape and higher order spacetime metric corrections. 
Then we use the theoretical model to fit synthetic PA data  to estimate how well the NS parameters can be constrained and to see how the results differ if an incorrect shape of the star is assumed.
We conclude in Sect.~\ref{sec:conclusions}.

\section{Theory}\label{sec:theory}

\subsection{NS shape and mass}\label{sec:geometry}

We consider a NS of mass $M$ rotating at a high frequency $\nu_* >300$~Hz, at which the stellar shape flattens significantly \citep{CMLC07,AGM14}. 
The shape of the star is given by the Schwarzschild coordinate $R(\theta)$, which is a function of the co-latitude.
We use one of the recent rapidly rotating NS shape models presented by \citet{AGM14}.
According to the model, the star shape is given by the approximation 
\be
R(\theta)=\req\left(1+o_2(x,\bar\Omega)\cos^2\theta\right), 
\ee 
where the flattening parameter, which defines the difference between the equatorial radius $\req$ and the polar radius $\rpol$,
is given in the first-order approximation as
\be
o_2(x,\bar\Omega)=\frac{\rpol-\req}{\req}\approx\bar\Omega^2(1.030x-0.788). 
\ee
Here 
\be
x=\frac{r_{\rm s}}{2\req}
\ee
is the dimensionless compactness parameter, $r_{\rm s}=2GM/c^2$ is the Schwarzschild radius of the NS, and 
\be
\bar\Omega=
2 \pi \nu_* \left(\frac{\req^3}{GM}\right)^{\frac12}
\ee
is the dimensionless circular rotation frequency. 
This approximation gives the shape of the star with better than 1\% accuracy when $\bar\Omega^2<0.1$, which is fulfilled for most of the realistic equations of state for NSs with $\nu_*<700$ Hz and $M>M_{\odot}$. 

\begin{figure}
\centering
\resizebox{0.9\hsize}{!}{\includegraphics{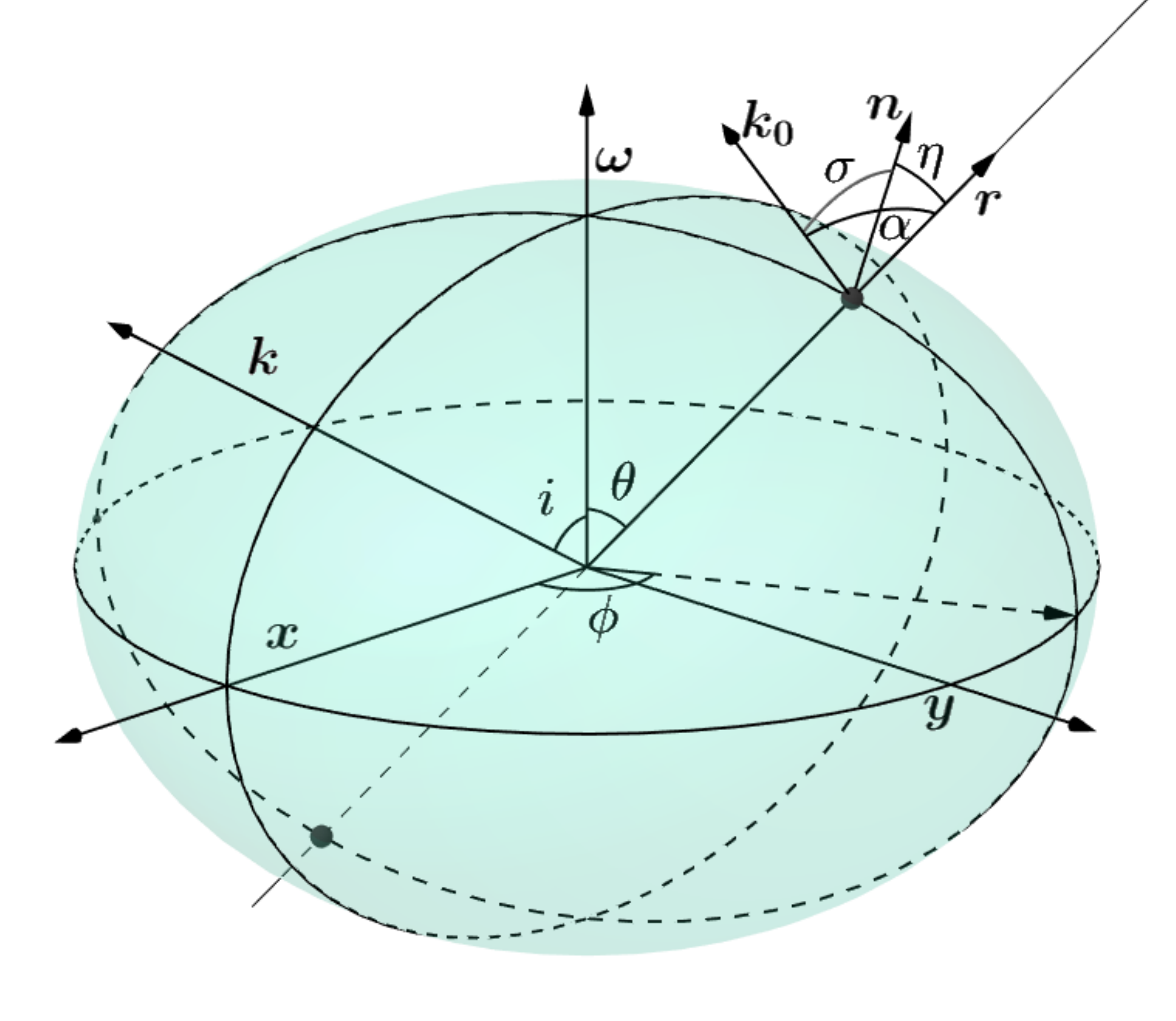}}
\caption{Geometry of the oblate star and the lab frame coordinate system $\bm{xyz}$. A spot at the NS surface at co-latitude $\theta$ has a rotation phase $\phi$ from the observer vector $\unit{k}$ around rotational axis $\bm z$. Three unit vectors are defined for the spot: $\unit{n}$ is normal to the surface, $\unit{r}$ is the radius vector, and $\unit{k}_0$ is the direction of emitted photon propagation.}
\label{fig:geometry}
\end{figure}

When comparing results for oblate and spherical stars, we also account for the difference in mass and the equatorial radius of the same NS due to its rapid rotation. 
For this purpose, following \citet{SP20}, we relate the gravitational mass $M$ and equatorial radius $\req$ of a rotating NS to the mass $M_0$ and radius $R_0$ of a spherical non-rotating NS of the same baryonic mass,
\be \label{eq:massprime}
M=M_0\left(a_0  + \frac{a_1}{1.1-\bar \nu}+ a_2\,\bar \nu^2\right) 
\ee
and
\be \label{eq:RRe}
 \req =R_0\,\left(0.9766+\frac{0.025}{1.07-\bar \nu}+\,0.07\,M_{1.4}^{3/2}\,\bar \nu^2\right) ,
\ee
where the coefficients are $a_0=1-a_1/1.1$, $a_1=0.001 M_{1.4}^{3/2}$, and $a_2=10 a_1$.
Here $M_{1.4}=M_0/1.4~\msun$ and $\bar \nu = {\nu_*}/{\nu_{\rm cr}}$ is the relative rotation frequency in units of the critical rotation frequency 
\be \label{eq:critical_nu}
\nu_{\rm cr} = 1278\,M_{1.4}^{1/2}\, \left(\frac{10~{\rm km}}{R_{0}}\right)^{3/2}\ \ \mathrm{Hz}.
\ee

The dimensionless speed (in units of speed of light $c$) of the NS surface at a given co-latitude as measured by a static observer is
\be \label{eq:betatheta}
\beta(\theta)=\frac{v}{c} = \frac{2 \pi R(\theta) \sin\theta}{\sqrt{1-u(\theta)}} \frac{\nu_*}{c},
\ee
where we have corrected the observed pulsar frequency for gravitational redshift, which can be expressed as 
\be
\frac{1}{\sqrt{1-u(\theta)}}=   \left(1-\frac{r_{\rm s}}{R(\theta)} \right)^{-1/2}= 1+z(\theta).
\ee
The corresponding Lorentz factor at a given $\theta$ is 
\be\label{eq:Lorentz-factor}
\gamma(\theta)=\frac1{\sqrt{1-\beta^2(\theta)}}.
\ee

The angle $\eta$ between the local normal to the NS surface and the radial direction depends only on latitude.
The values of trigonometric functions for this angle can be represented as 
\be
\cos\eta=\frac1{\sqrt{ 1+ f^2(\theta) }}
\quad\mbox{and} \quad
\sin\eta=\frac{f(\theta)}{\sqrt{ 1+ f^2(\theta) }},
\ee
where the function $f(\theta)$ can be expressed in terms of function $R(\theta)$ and its derivative
\be\label{eq:shape}
        f(\theta)= \frac{1+z(\theta)}R\frac{\diff R}{\diff\theta}.
\ee

\subsection{Vectors and angles}\label{sec:angles}

Let us next recall the vectors and the angles between them in the OS approximation. 
The geometry of the star is the same as in \citet{SNP18}, and it is shown in Fig. \ref{fig:geometry}.
The axis of pulsar rotation coincides with the symmetry axis, and we choose it to be the applicate axis  $\bm{z}$ in our fixed laboratory frame. 
The unit vector $\unit{k}$ pointing to the direction of the observer from the star center lies in the $\bm{xz}$-plane and makes angle $i$ with the rotation axis: 
\be
        \unit{k}=(\sin i, 0, \cos i).
\ee
The $\bm y$-axis completes the right-handed Cartesian system.

We consider first a small hotspot being placed on the star surface and a constant co-latitude $\theta$. 
For larger spots we integrate observed Stokes parameters over the spot area. 
The phase of the pulsar $\phi$ is the angle measured counterclockwise around the $\bm z$-axis. 
 We define the phase to be zero when the spot (or its center for large circular spots) passes through the $\bm{xz}$-plane. 
When we consider more than one spot we identify the spot closest to the observer as the main emission region, which also defines the phase.

For every moment of time we define three local unit vectors on the surface related to the spot.
They are the radius vector $\unit{r}$ of the spot, the vector normal to the spot $\unit{n}$, and the vector $\unit{k}_0$ denoting the initial direction of light propagation emitted so that a photon eventually arrives at the observer. 
Correspondingly, we have three angles between these vectors, so that $\eta$ is between $\unit{n}$ and $\unit{r}$, $\alpha$ is between $\unit{r}$ and $\unit{k}_0$, and $\sigma$ is between $\unit{n}$ and $\unit{k}_0$.
The applied coordinate system, unit vectors, and related angles are visualized in Fig. \ref{fig:geometry}.

Each vector can be represented by a set of coordinates in the lab system. Thus, the radius-vector $\unit{r}$ is 
\be 
\unit{r}=(\sin\theta\cos\phi,\sin\theta\sin\phi,\cos\theta),
\ee
and the normal is 
\be 
\unit{n}=(\sin\lambda\cos\phi,\sin\lambda\sin\phi,\cos\lambda),
\ee
where $\lambda = \theta-\eta$ is the angle between the normal vector and the spin axis.
The vector $\unit{k}_0$ lying between $\unit{r}$ and $\unit{k}$ is expressed as
\be\label{eq:kzero}
        \unit{k}_0=\frac{1}{\sin\psi}\left( \sin\alpha\  \unit{k}+\sin(\psi-\alpha)\ \unit{r}\right) ,  
\ee
where the angle $\psi$ between $\unit{r}$ and $\unit{k}$ is determined by the scalar product of these vectors: 
\be\label{eq:cospsi}
        \cos \psi = \unit{r} \cdot \unit{k} = 
         \cos i \cos\theta + \sin i \sin\theta\cos\phi .
\ee
The relation between angle $\alpha$ and $\psi$ depending on the radius of the emission point $R(\theta)$ is given by the light bending integral \citep{PFC83,B02}, which can be computed either exactly as described in \citet{SNP18}, for example, or via the approximate analytical formulae in \citet{poutanen20a}. 
The angle $\sigma$ can be expressed as \citep{MLC07,SP20}
\beq \label{eq:cos_sigma}
\cos\sigma & = & \unit{n} \cdot \unit{k}_0 = \cos \alpha  \cos \eta  \nonumber \\ 
 & + & \frac{\sin \alpha}{\sin \psi} \sin \eta\ (\cos i \sin \theta - \sin i \cos \theta \cos \phi).
\eeq 

Because the star is rotating, the spot is moving in the lab frame with dimensionless velocity $\bm{\beta}=\unit{\beta} \beta$,
which we decompose into the product of the unit velocity vector 
\be\label{betahat}
        \unit{\beta} =(-\sin\phi,\cos\phi,0)
\ee
and the dimensionless velocity $\beta$ given by Eq.\,\eqref{eq:betatheta}.

\subsection{Observed flux and polarization degree}\label{sec:flux_pd}

The flux observed at energy $E$ from a small spot of area $dS'$ (measured in the comoving frame) can be obtained from the following equation \citep{PB06,MLC07,SNP18,SP20}:
\be \label{eq:fluxspot2}
d F_{E}=(1-u)^{3/2} \delta^{4} I'(E',\mu) \cos\sigma\ 
{\cal D} \frac{\diff S'}{D^2} .
\ee
Here $u$ is the compactness at the emission point; 
\be
        \delta=\frac{1}{\gamma(1-\beta \cos\xi)} 
\ee
is the Doppler-factor, where $\xi$ is the angle between the spot velocity vector and the photon direction, 
\be \label{eq:cosxi}
\cos\xi = \unit{\beta} \cdot \unit{k}_0 = -  \frac{\sin\alpha}{\sin\psi} \sin i\ \sin\phi\ ;
\ee
$\mu$ denotes the cosine of the inclination $\sigma'$  of the line of sight to the normal in the co-rotating reference frame expressed as \citep{PG03,PB06}
\be
\mu=\cos\sigma' = \delta\cos{\sigma};  
\ee
and $I'(E',\mu)$ is the spectral intensity measured in the comoving frame at the zenith angle $\sigma'$. 
Furthermore, 
\be
{\cal D}=\frac{1}{1-u} \frac{\diff \cos\alpha}{\diff\cos\psi}
\ee
is the lensing factor and $D$ is the distance to the star (we use $D=1$ since it does not affect the PA profile).
We note that $D \gg R$ to ensure the validity of the presented equations because we implicitly assume that the observer's image plane is at infinity.

The focus of this paper is the effects of NS oblateness on the PA, which means that the choice of the spectral model is not very important. 
We approximate the spectrum of the emitted photons in the local corotating frame at energy $E'$ with the Planck function $B_{E'}(T_{\rm bb})$. 
However, the dependence of intensity on cosine of the zenith angle $\mu=\cos\sigma'$ is taken from that expected for X-ray bursting millisecond pulsars \citep{1947ApJ...105..435C,Cha60,sob49,Sob63,SPW12}:
\be \label{eq:beaming}
I(E',\mu) = B_{E'}(T_{\rm bb})\ (0.421+0.868\,\mu).
\ee
Because the PD does not change along photon trajectory, the observed PD from a small spot is fully determined by the PD at the zenith angle $\sigma'$ as measured in the comoving frame. 
For the simulations below we assume that  the angular dependency of the PD corresponds to that of the electron-scattering dominating atmosphere, which we approximate as \citep{VP04}
\be \label{eq:pol_deg}
  P (\mu)=- \frac{1-\mu}{1+3.582 \mu} 11.71 \%.
\ee
The negative value for the PD means that the dominant direction of the oscillation of the electric vector is perpendicular to the plane formed by the local normal and the photon momentum.

For a rapidly rotating NS, time delays of photon propagation around the star have to be accounted for. 
The observed phase (i.e., the phase when photon arrives to the observer) differs from the rotation (emission) phase $\phi$, 
\be 
\varphi = \phi + 2 \pi \nu_* \Delta t,
\ee 
where $\Delta t$ is the time delay calculated as explained  in \citet{SNP18}, among others. 
However, unlike there, we use the radius at the center of the main spot $R(\theta)$ as a reference radius, and a reference emission angle is chosen to correspond to the emitted photon direction from that point at $\phi=0$.
By this definition, the time delay $\Delta t$ is zero for a single small spot at $\phi=0$, and thus same for both oblate and spherical stars at that phase.

\subsection{Polarization angle}\label{sec:pa_eqs}

Let us now present derivation of the PA, which differs from that presented for spherical stars in the previous papers \citep{VP04,poutanen20b}. 
There are several steps that need to be  made in order to transform the Stokes vector from the co-rotating frame of the star to the sky plane.
Let us first define the main polarization basis, in the sky plane, formed by $\unit{z}$ and $\unit{k}$, 
\be 
\label{eq:mbasis}
\unit{e}_1^{\mathrm{m}} = \frac{\unit{z}-\cos{i} \  \unit{k}}{\sin{i}},\qquad 
\unit{e}_2^{\mathrm{m}} = \frac{\unit{k} \times \unit{z}}{\sin{i}}.
\ee
The vector $\unit{e}_1^{\mathrm{m}}$ is colinear to the projection of the star's rotation axis at the sky plane and the second vector $\unit{e}_2^{\mathrm{m}}$ is perpendicular to the first one.
This basis is fixed in the lab frame because the pulsar rotation axis (along the $\unit{\omega}$) is very stable and the line of sight (along the $\unit{k}$) does not change during the observation.
The polarization (pseudo-)vector is observed in this basis.
The PA defines the direction of the polarization vector in a basis, and it is customary to measure it in the counterclockwise direction from the first basis vector $\unit{e}_1^{\mathrm{m}}$.

Next let us consider the basis formed by $\unit{r}$ and $\unit{k}$:
\be\label{eq:rbasis}
\unit{e}^{\psi}_1 = \frac{\unit{r}-\cos{\psi}\ \unit{k}}{\sin{\psi}},\qquad 
\unit{e}^{\psi}_2 = \frac{\unit{k} \times \unit{r}}{\sin{\psi}}.
\ee
This basis is defined in the same plane as the previous one, as it is also perpendicular to $\unit{k}$. 
However, this basis is rotated relative to the main one. 
The angle between these two bases we denote by $\chi_0$, and the corresponding trigonometric functions are
\be
\cos{\chi_0}=\unit{e}_1^{\mathrm{m}} \cdot \unit{e}^{\psi}_1 = \frac{\sin{i}\cos{\theta}-\cos{i} \sin{\theta}\cos{\phi}}{\sin{\psi}} 
\ee 
and 
\be \label{eq:chi0}
\sin{\chi_0}= - \unit{e}_1^{\mathrm{m}} \cdot \unit{e}^{\psi}_2 = \unit{e}_2^{\mathrm{m}} \cdot \unit{e}^{\psi}_1 = - \frac{\sin{\theta}\sin{\phi}}{\sin{\psi}} .
\ee
Thus, we get the standard expression for the PA of the rotating vector model (RVM; \citealt{RC69}):
\be \label{eq:pa_rvm}
\tan\chi_0= \frac{-\sin \theta\ \sin \phi}
{\sin i\ \cos \theta  - \cos i\ \sin \theta\  \cos \phi }.
\ee
For the spherical star and slow rotation this is sufficient; for the  oblate NS shape and rapid
rotation  additional corrections are needed.

To take into consideration the surface curvature we define another basis formed by the radius vector $\unit{r}$ and the direction of the light propagation $\unit{k}_0$ near the NS surface:
\be\label{eq:basis}
\unit{e}^{\alpha}_1 = \frac{\unit{r}-\cos{\alpha}\  \unit{k}_0}{\sin{\alpha}},\qquad 
\unit{e}^{\alpha}_2 = \frac{\unit{k}_0 \times \unit{r}}{\sin{\alpha}}=\unit{e}^{\psi}_2.
\ee
This basis describes the plane perpendicular to the initial photon direction close to the NS surface.
Because in the Schwarzschild metric the light trajectories are planar, vectors $\unit{e}^{\alpha}_2$ and $\unit{e}^{\psi}_2$ coincide. 
The PA measured in this basis near the surface is the same as that measured by the instrument in the basis denoted by upper index $\psi$ in Eq.\,\eqref{eq:rbasis} because the polarization vector is parallel-transported along photon trajectory. 
In addition, we introduce the basis associated with the local normal vector $\unit{n}$: 
\be\label{eq:nbasis}
\unit{e}_1^{\sigma} = \frac{\unit{n}-\cos{\sigma}\  \unit{k}_0}{\sin{\sigma}},\qquad 
\unit{e}_2^{\sigma} = \frac{\unit{k}_0 \times \unit{n}}{\sin{\sigma}}.
\ee
This basis   lies in the same plane as the previous one, perpendicular to $\unit{k}_0$.
The angle between them, $\chi_1$, can be computed from 
\be
\cos{\chi_1}=\unit{e}_1^{\sigma} \cdot \unit{e}^{\alpha}_1 = \frac{\cos\eta-\cos\alpha\cos\sigma}{\sin{\alpha}\sin\sigma} 
\ee
and 
\begin{align}
\sin{\chi_1} &= \unit{e}^{\alpha}_2 \cdot \unit{e}^{\sigma}_1 = \frac{\unit{n} \cdot (\unit{k}_0\times\unit{r} )}{\sin\alpha\sin\sigma}
= \frac{\unit{k}_0 \cdot (\unit{r}\times\unit{n} )}{\sin\alpha\sin\sigma} \nonumber \\
&= \frac{\sin\alpha}{\sin\psi} \frac{ \unit{k} \cdot (\unit{r}\times\unit{n} )}{\sin\alpha\sin\sigma}
= \frac{ \sin\eta\sin i \sin\phi}{\sin\psi\sin\sigma}.
\end{align}
Thus, we get 
\be \label{eq:chi1}
\tan{\chi_1}= - \frac{\sin\eta\cos\xi}{\cos\eta-\cos\alpha\cos\sigma}.
\ee
If the star is spherical (i.e., $\eta=0$), angle $\chi_1$ is exactly zero.

The last step is to take into account the positional vector rotation due to the relativistic motion of the NS surface. 
The relativistic correction of the polarization vector we denote by $\chi'$. 
It can be shown that Lorentz transformation of the polarization basis vector $\unit{e}_1^{\sigma}$ is \citep[see, e.g.,][]{NP93} 
\be\label{eq:primebasis}
\unit{e}_1' = \frac{\unit{n}-\mu (\delta\unit{k}_0-\gamma\bm{\beta})}{\sqrt{1-\mu^2 } } .
\ee
 
The expressions for the trigonometric functions of relativistic rotation angle of the polarization plane are 
\begin{align}\label{eq:coschiprime}
\cos\chi'&=\unit{e}_1'\cdot \unit{e}_1^{\sigma} =
\frac{1-\cos^2\sigma- \delta  \gamma \cos^2 \sigma  \beta\cos\xi }{\sin{\sigma}\sqrt{1-\mu^2} } \nonumber  \\
&=\frac{\sin^2\sigma - \beta\cos\xi }{\sin{\sigma}\sqrt{1-\mu^2} (1-\beta\cos\xi) } 
\end{align}     
and
\begin{align}\label{eq:sinchi}
\sin{\chi'}&=\unit{e}_1'\cdot \unit{e}_2^{\sigma} =
\frac{\mu\gamma\beta}{\sin{\sigma}\sqrt{1-\mu^2} } \unit{\beta} \cdot(\unit{k}_0 \times \unit{n}) \nonumber
\\
&=\frac{\beta\cos\sigma }{\sin{\sigma}\sqrt{1-\mu^2}  (1-\beta\cos\xi)} \unit{k}_0  \cdot(\unit{n}\times \unit{\beta}).
\end{align}
The vector product $\unit{n}\times \unit{\beta}$ is essentially the meridional vector pointed towards the north pole:
\be
\unit{m} = \unit{n}\times \unit{\beta} = (- \cos \lambda \cos \phi, -\cos \lambda \sin \phi, \sin \lambda ) .
\ee
Then using Eq. \eqref{eq:kzero} we can express the scalar triple product as
\be\label{eq:tripleproductpherical}
\unit{k}_0 \cdot (\unit{n}\times\unit{\beta} )=\unit{k}_0 \cdot \unit{m} 
= \frac1{\sin\psi}\left(\sin \alpha \cos \zeta - \sin(\psi-\alpha)\sin\eta \right),
\ee
where $\zeta$ is the angle the vector $\unit{k}$ makes with the meridian 
\begin{equation}\label{eq:zeta_defined}
\cos \zeta = \unit{m} \cdot \unit{k} = \cos i \sin \lambda - \sin i \cos \lambda \cos \phi.
\end{equation}
Thus, we get 
\be\label{eq:sinchiprime}
\sin\chi'={\beta \cos\sigma }
\frac{
\cos\zeta \sin\alpha-\sin(\psi-\alpha) \sin\eta}
{\sin\psi\sin{\sigma}\sqrt{1-\mu^2}  (1- \beta \cos\xi) }.
\ee  
Combing Eqs.~\eqref{eq:cosxi}, \eqref{eq:coschiprime}, and \eqref{eq:sinchiprime} we get 
\be\label{eq:chiprime}
\tan\chi' = \beta \cos \sigma\frac{\cos\zeta -\sin \eta  \frac{\sin (\psi-\alpha)}{\sin \alpha}}{\frac{\sin \psi}{\sin \alpha}\sin^{2}\sigma + \beta \sin i \sin \phi}.
\ee
For the case of a spherical star we can set $\eta=0$ and recover
\be\label{eq:chiprime_sphere}
\tan\chi'_{\mathrm{sph}} = 
\beta \cos \alpha \frac{\cos i \sin \theta - \sin i \cos \theta \cos \phi}{\sin \psi\sin\alpha + \beta \sin i \sin \phi },
\ee
which coincides with Eq. (29) from \citet{VP04} and derived by \citet{poutanen20b}.
An alternative derivation of Eq. \eqref{eq:chiprime} is  presented in Appendix \ref{sec:appendix}. 

The total PA for each spot is then obtained by summing the angles between the intermediate bases since they are constant for parallel transport along the photon trajectory: 
\be\label{eq:chi_tot}
\chi_{\mathrm{obl}}=\chi_0+\chi_1+ \chi'.
\ee
Similarly, for the spherical case this reads
\be\label{eq:chi_tot_sph}
\chi_{\mathrm{sph}}=\chi_0+\chi'_{\mathrm{sph}}.
\ee

\subsection{Observed Stokes parameters}\label{sec:Stokes} 

Let us finally describe how to obtain the final Stokes parameters and the total PA. 
Because we do not consider circular polarization, the Stokes vector that describes observed radiation from an infinitely small spot of area $\diff S'$ contains three components $(\diff F_{\rm I},\diff F_{\rm Q},\diff F_{\rm U})$. 
Without losing the generality, we assume that the PA equals zero in the polarization basis defined in the co-moving frame of the spot because we expect azimuthal symmetry of the radiation in that frame. 
For the case of the electron-scattering dominated atmosphere the dominant direction of oscillations of the electric vector is perpendicular to the normal;  we  can still take PA=0\degr, but assign a negative value for the PD $P$.
Because the PD does not change along photon trajectory, the observed Stokes vector is fully defined by the flux $\diff F_{\rm I}$, PD $P$, and   PA $\chi$: 
\be \label{eq:Stokes1}
\left( \begin{array}{c}
\diff F_{\rm I} \\
\diff F_{\rm Q} \\ 
\diff F_{\rm U} \\
\end{array} \right) 
= \diff F_{\rm I} \left( \begin{array}{c}
1 \\
P \cos{2\chi} \\ 
P \sin{2\chi} \\
\end{array} \right) .
\ee
 
When the spots have a finite size, we need to integrate over the spot area in the comoving frame expressed as \citep{NP18,SNP18,Bogdanov19L26,SP20}
\be 
\diff S'=\gamma R^2(\theta)\ \diff\cos\theta\ \diff\phi'/\cos\eta .
\ee 
For each element of the spot we need to compute a separate light curve as a function of observed phase for the flux, PD, and PA, accounting for different time delays, and to sum them up. 
The total Stokes vector (denoted by the index {``$\mathrm{tot}$''}) is obtained by integrating the Stokes vectors given by Eq.\,\eqref{eq:Stokes1} over the NS surface.
The corresponding total PD is then
\be
P^{\mathrm{tot}}=\frac{\sqrt{(\ftot_{\rm Q})^2+(\ftot_{\rm U})^2}}{\ftot_{\rm I}},
\ee
where $\ftot_{\rm I}$ represents the total flux, and the PA of the total Stokes vector $\chi^{\mathrm{tot}}$ can be computed from
\be
\cos{2\chi^{\mathrm{tot}}}=\frac{\ftot_{\rm Q}}{\ftot_{\rm I}}, \qquad
\sin{2\chi^{\mathrm{tot}}}=\frac{\ftot_{\rm U}}{\ftot_{\rm I}}.
\ee

\begin{figure}
\resizebox{0.9\hsize}{!}{\includegraphics{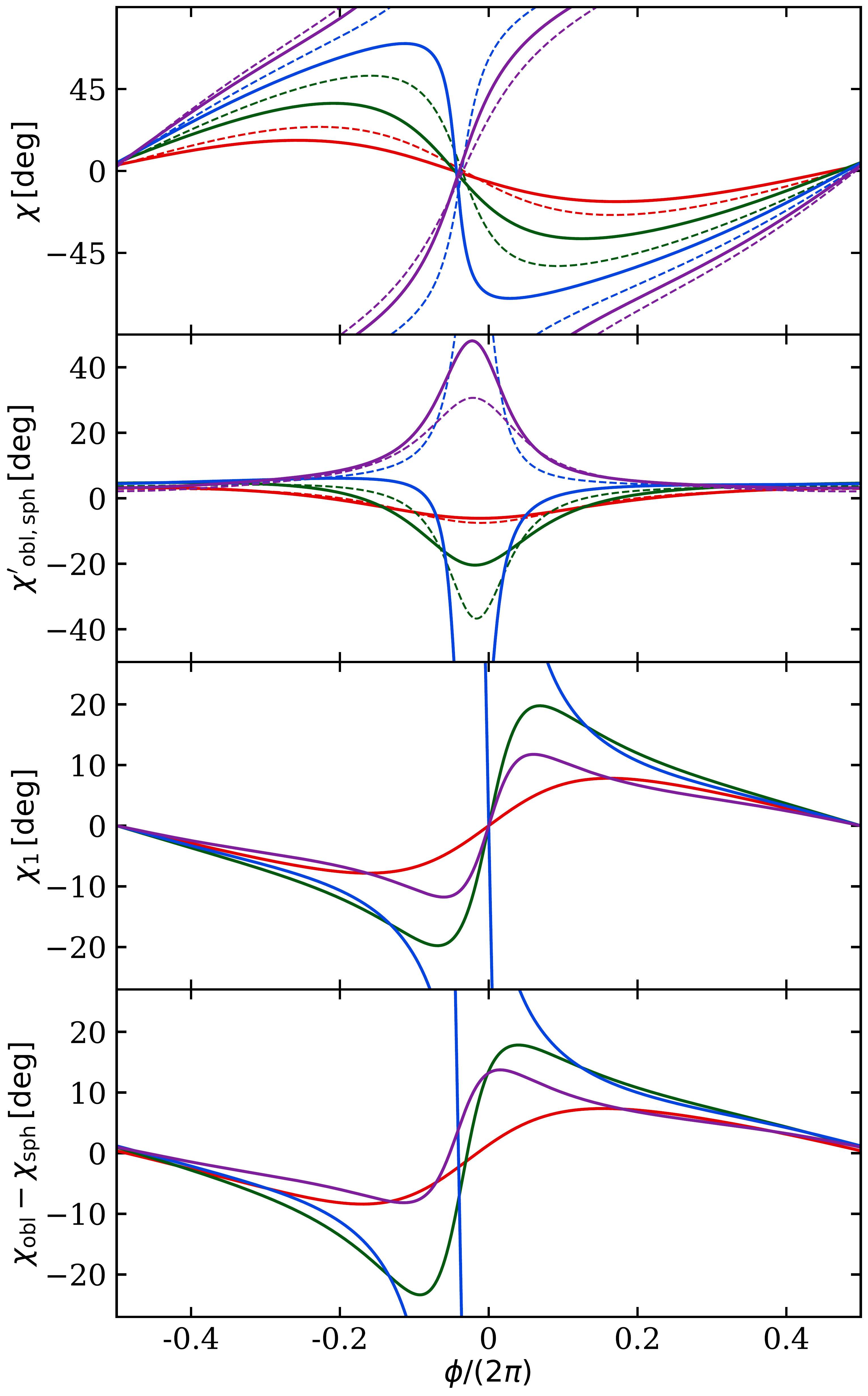}}
\caption{\textit{Top panel}: 
Polarization angles  computed for the spherical (dashed) and the oblate (solid) star as a function of the rotation phase $\phi$ for different spot co-latitudes $\theta$=15\degr\ (red), 30\degr\ (green), 45\degr\ (blue), and 60\degr\ (purple). 
\textit{Second panel}: 
Relativistic rotation for the spherical (dashed) and the oblate (solid) stars. 
\textit{Third panel}: 
Component $\chi_1$ for the oblate star. 
\textit{Bottom panel}: 
 Difference between the PAs for the oblate and spherical stars. 
Parameters of the non-rotating NS are $R_0 = 12$~km and $M_0 = 1.4~\msun$, while in the oblate case of a NS rotating at a rate of $\nu_* = 600$ Hz, the mass and radius are computed according to Eqs.~\eqref{eq:RRe} and \eqref{eq:massprime}. 
The inclination is $i = 40\degr$.
}
\label{fig:diff_theta}
\end{figure}

\begin{figure}
\resizebox{0.9\hsize}{!}{\includegraphics{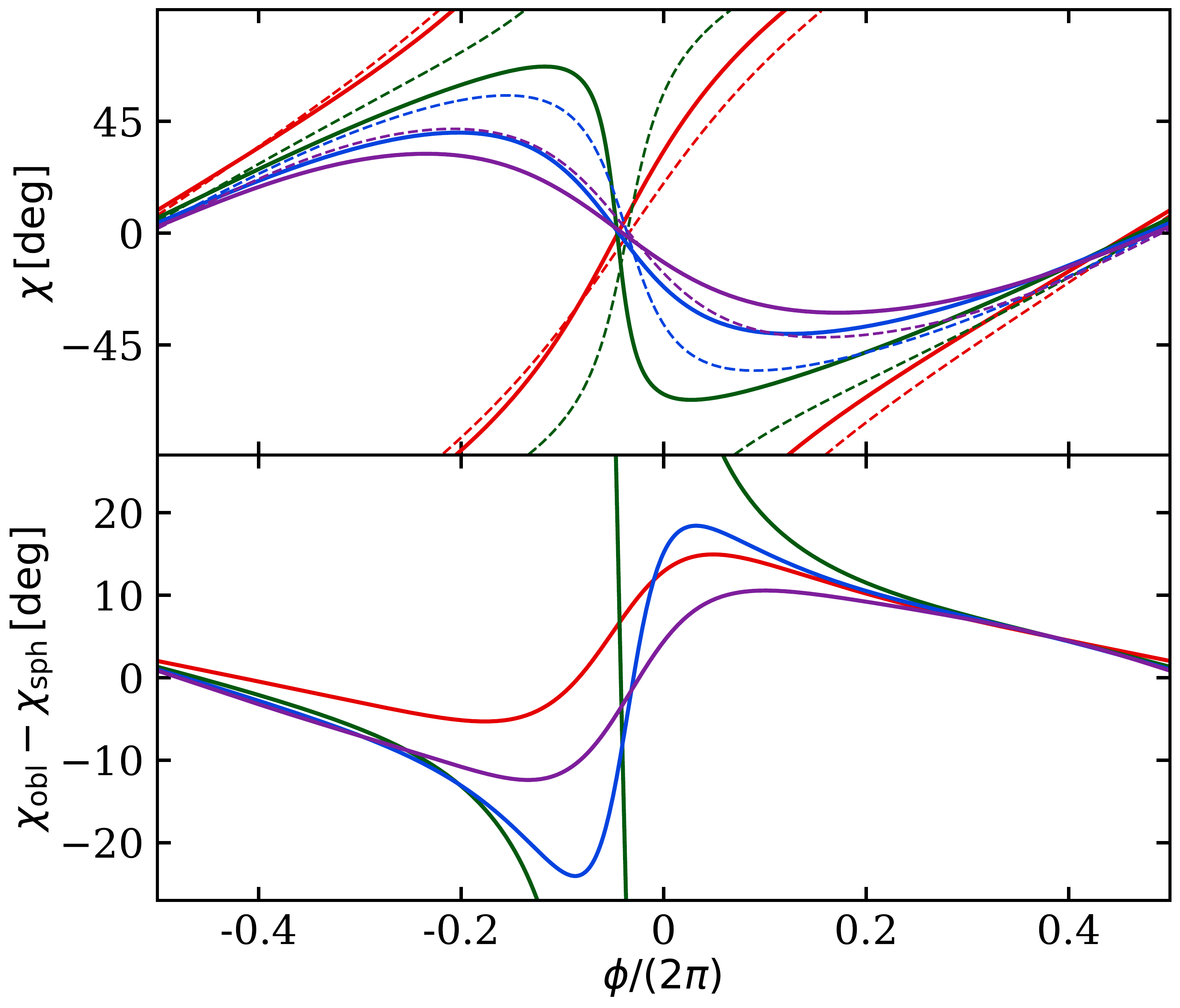}}
\caption{Same as Fig.~\ref{fig:diff_theta} (only the top and bottom panels), but for different inclinations $i$=15\degr\ (red), 30\degr\ (green), 45\degr\ (blue), and 60\degr\ (purple). 
Here the spot co-latitude is $\theta = 35\degr$.
}
\label{fig:diff_incl}
\end{figure}

\begin{figure}
\resizebox{0.9\hsize}{!}{\includegraphics{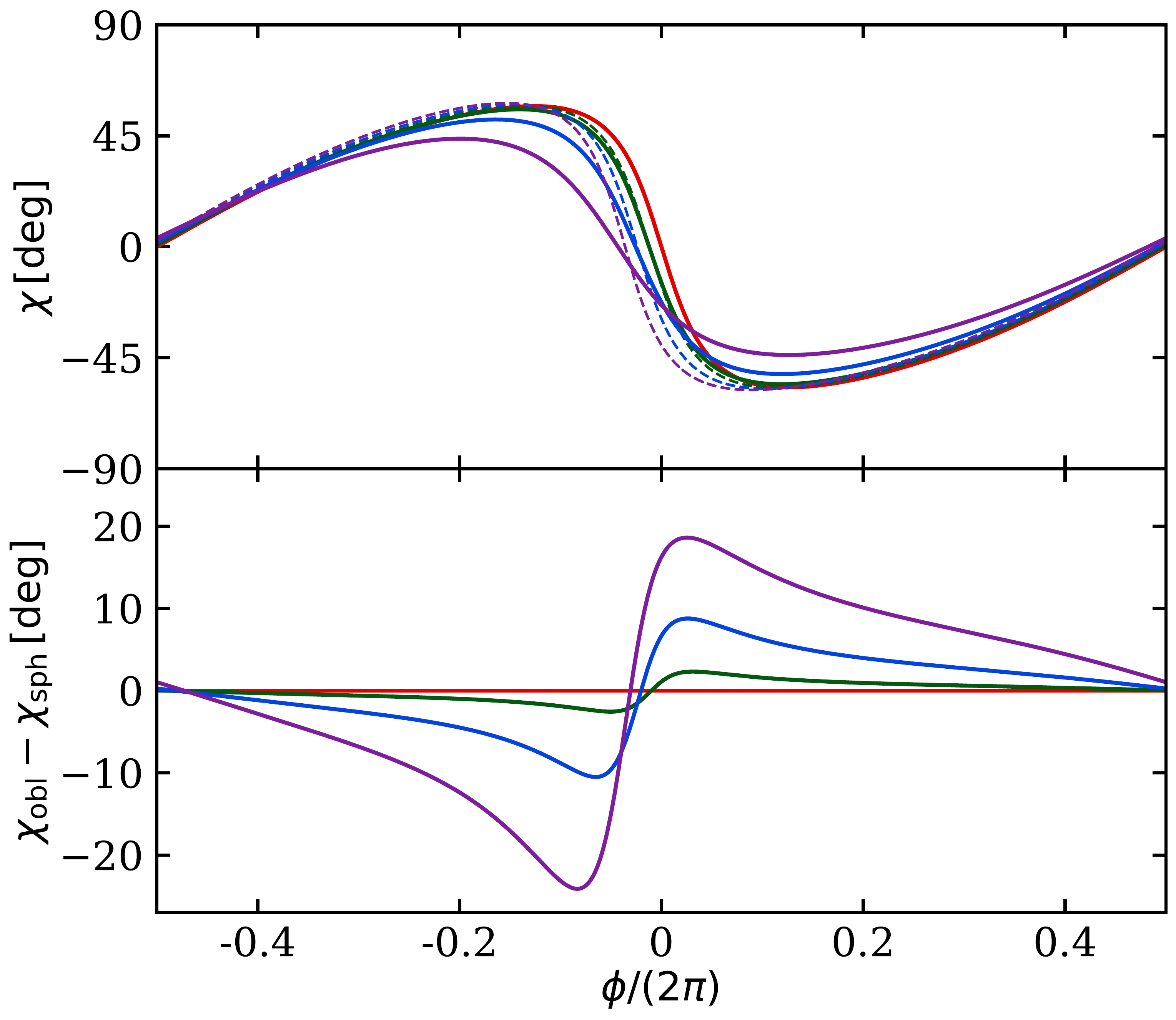}}
\caption{Same as Fig.~\ref{fig:diff_incl}, but for different frequencies $\nu_* = 1$ (red), 200 (green), 400 (blue), and 600~Hz (purple).
Here the spot co-latitude is $\theta = 40\degr$ and inclination is $i=50\degr$.  
}
\label{fig:diff_nu}
\end{figure}

\begin{figure}
\resizebox{0.9\hsize}{!}{\includegraphics{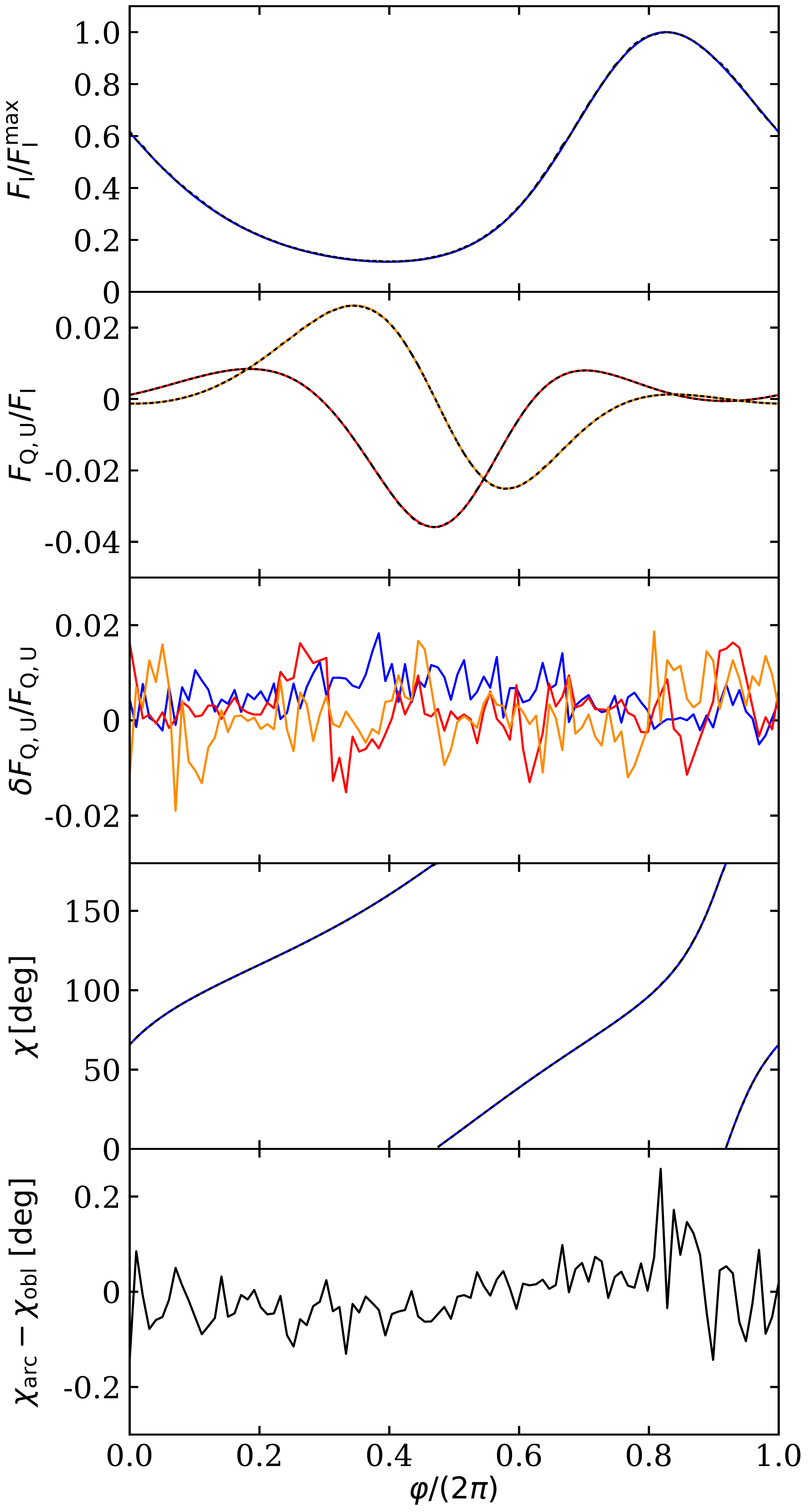}}
\caption{
Comparison of Stokes parameters and PA profiles for our analytical OS approximation against the numerical OS approximation calculated with \textsc{arcmancer}. 
The three upper panels show the flux (blue curve), the two Stokes parameters ($F_{\mathrm{Q}}$ in red and $F_{\mathrm{U}}$ in orange), and the residuals in all three Stokes parameters. 
All the corresponding results obtained using \textsc{arcmancer} are shown with black dashed curves that overlap almost exactly the colored lines. 
The two lower panels show the PA profile and the residuals between the results of the two codes.
Our results are shown by a blue curve and the results of \textsc{arcmancer} with an almost exactly overlapping black curve.
The profiles are computed for one spot at a NS with $\nu_* = 600$~Hz, $\req = 12$~km, $M = 1.4~\msun$, $\theta = 60\degr$, $i = 40\degr$, $\rho = 10 \degr$, $T_{\mathrm{bb}}=1$ keV, and $E = 5$~keV. 
} 
\label{fig:arc_obl_obl}
\end{figure}

\section{Results}\label{sec:results}

\subsection{Comparing PA for oblate and spherical stars}

We begin by comparing the analytic expressions for the PA in Eqs.\,\eqref{eq:chi_tot} and \eqref{eq:chi_tot_sph} in Figs.~\ref{fig:diff_theta}--\ref{fig:diff_nu}, for different co-latitudes, inclinations, and spin frequencies, and show them as functions of the rotation phase $\phi$. 
The resulting pulse profiles look very similar if we present them as functions of the observed phase $\varphi$. 
For a spherical star we took the radius $R_0=12$\,km and mass $M_0=1.4\msun$. 
For an oblate star we computed the gravitating mass and equatorial radius using Eqs.\,\eqref{eq:massprime} and \eqref{eq:RRe}.

From Figs.~\ref{fig:diff_theta} and \ref{fig:diff_incl} we see that deviation between the spherical and oblate PA profiles (for $\nu_*=600$ Hz) depends strongly on the combination of the angles $i$ and $\theta$ and on the phase $\phi$. 
The difference is the smallest (less than about 10 degrees) when either $\theta \ll i$ or $\theta \gg i$. 
However, for $\theta \approx i$ we see a large deviation in the vicinity of $\phi\approx 0$, where the total angle $\chi$ changes the sign. 
In these situations the dashed and solid lines (for spherical and oblate stars, respectively) have opposite monotonicities. 
The explanation is that for the oblate case the condition of $\lambda  = \theta -\eta < i$ can be satisfied even though $\theta>i$ for a spherical star. 
For example, at a spin rate of $\nu_* = 600$ Hz, the angle $\eta$ can be as large as $7\degr$ at $\theta = 45\degr$.
Nonetheless, we note that the PD is also the smallest at $\phi \approx 0$, since the direction towards the observer is closest to the local normal to the surface.
The small difference between angles $\theta$ and $i$ leads to the closest alignment between the normal and the trajectory of the light on these phases, and thus the high difference in the PA may still be difficult to observe.

\begin{figure}
\resizebox{0.9\hsize}{!}{\includegraphics{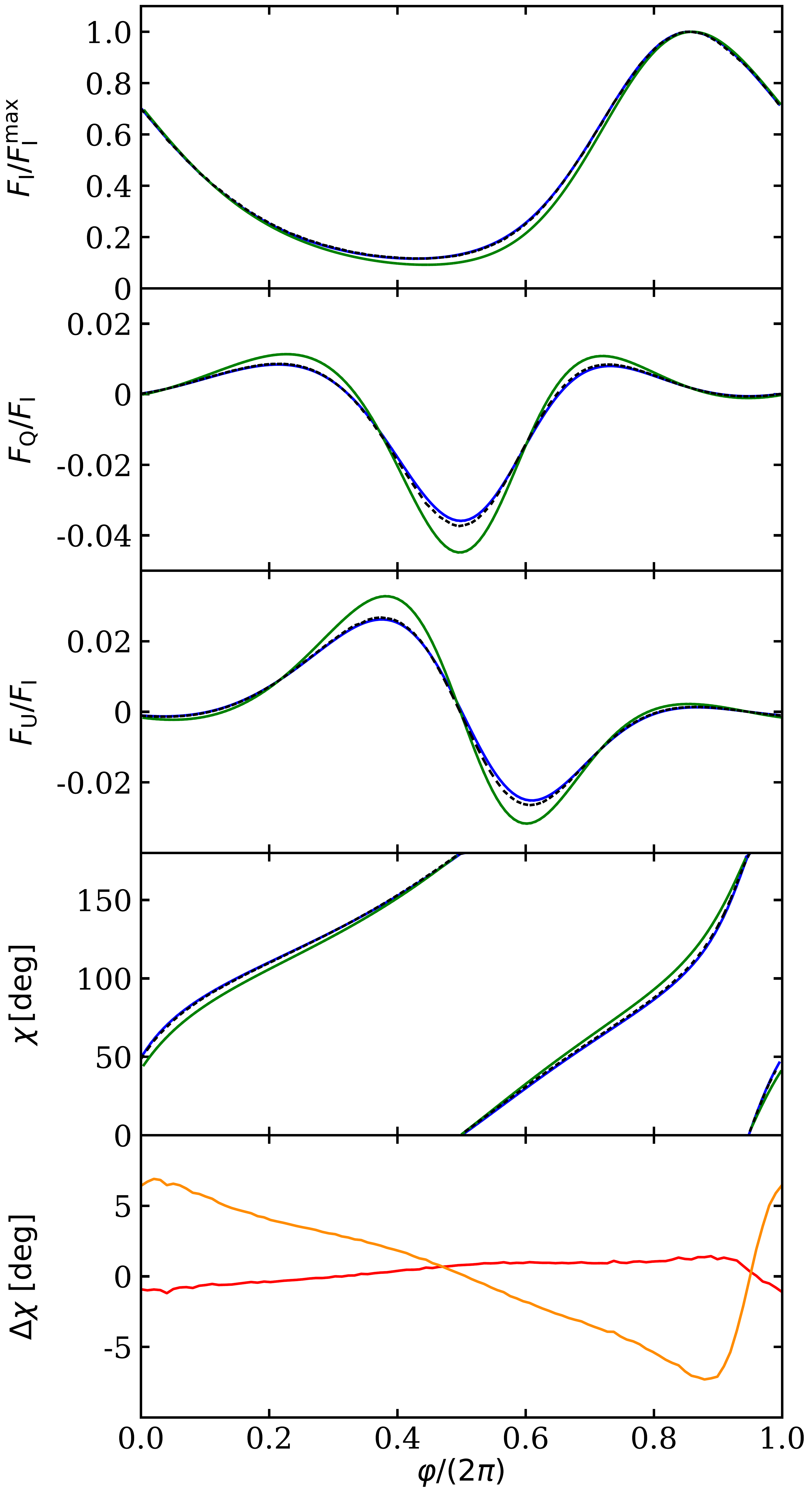}}
\caption{
Stokes parameters and PA profiles for our analytical OS model (blue curve), the exact result of \textsc{arcmancer} with AGM model (dashed black curve), and the S+D model from \citet{VP04} (green curve). 
The panels, from  top to  bottom, show the flux normalized to the maximum, the normalized Stokes parameters $F_{\mathrm{Q}}$ and $F_{\mathrm{U}}$, the PA, and the difference in PA ($\chi_{\mathrm{AGM}}-\chi_{\mathrm{OS}}$ in red and $\chi_{\mathrm{AGM}}-\chi_{\mathrm{S+D}}$ in orange), respectively. 
The blue and black curves almost completely overlap.
The parameters used to calculate the profiles are the same as in Fig. \ref{fig:arc_obl_obl}, except for a spherical star we use the mass and radius of the corresponding non-rotating NS of the same baryonic mass: $M_{0}=1.394 ~\msun$ and $R_{0} = 11.43$ km (see Sect.\,\ref{sec:geometry}).
}
\label{fig:arc_agm_obl_sph}
\end{figure}

The precise rotation phase $\phi$, where $\chi$ changes sign, is slightly smaller than zero because $\chi'$ reaches the extremum at an earlier phase where the denominator of Eq.\,\eqref{eq:chiprime} reaches the minimum. 
For the oblate case that phase is generally smaller than the corresponding phase for the spherical star, partly because $\chi'$ is generally larger than $\chi'_{\mathrm{sph}}$ close to $\phi=0$, as can be seen from Eqs.\,\eqref{eq:chiprime} and \eqref{eq:chiprime_sphere} and the second panel of Fig.~\ref{fig:diff_theta}. 
The angle $\chi_1$ is zero at $\phi = 0$. 
Around phases $\phi \approx \pm \pi/2$ the correction from $\chi_1$ dominates the differences in the total $\chi$. 
Thus, for instance, the PA for the oblate NS is closer to zero in comparison to its spherical counterpart for those cases when the PA does not go the full circle (i.e., when $i>\lambda$). 
This is a direct consequence of the oblate shape of the star: the angle $\chi_1$ defined by Eq.\,\eqref{eq:chi1} has a different sign than $\chi_0$ in Eq.\,\eqref{eq:chi0}. 
In addition, as can be seen from Fig. \ref{fig:diff_nu}, the difference in the total PA $\chi$ gets larger for faster rotating NSs, and it can reach ten degrees at about 400 Hz.  

\subsection{Comparison with polarized light tracing}

Let us now compare our results of polarized light curves to those computed with the general relativistic polarized radiative transfer code \textsc{arcmancer} \citep{PMN18}.
In that code, the PA profile is calculated numerically along the geodesics using a user-specified metric.
The method relies on numerically propagating a given set of coordinate vectors along the geodesic starting from the observer's image plane until the photon trajectory intersects with the NS surface.
This comparison was used to verify that our analytical formulae include all the necessary special and general relativistic sources of polarization plane rotation.

We found good agreement between the results of the codes when using the OS approximation for both codes as shown in Fig.\,\ref{fig:arc_obl_obl}.
The PA profiles and the Stokes parameters were calculated for one hotspot with $\nu_* = 600$~Hz, $\req = 12$~km, $M = 1.4~\msun$, $\theta = 60\degr$, $i = 40\degr$, angular radius of the spot $\rho = 10 \degr$, temperature of the spot $T_{\mathrm{bb}}=1$ keV, and photons observed at $E = 5$~keV energy. 
We observe a difference in the Stokes parameters of about 1\% and less than $0\fdg2$ difference in the PA.
However, we note that the polarization profiles of \textsc{arcmancer} are more sensitive to the distance of the image plane than in the case of flux \citep[comparisons between fluxes were made in][]{NP18}.
Therefore, to avoid systematic biases in the residuals, we had to increase the image plane distance from 100, which was suitable for light curve comparison, to more than $1000\,r_{\rm s}$. 

We also compared the results for the PA profiles computed in the OS approximation, spherical S+D approximation, and in the \citet{BI76} metric, 
around NS as described in \citet[][AGM]{AGM14}, where  the shape of the star and the space-time itself are no longer spherical  (see Fig.~\ref{fig:arc_agm_obl_sph}). 
We found that our analytical OS model matches  the AGM model computed with \textsc{arcmancer} much more closely than the spherical S+D model.  
Most of the polarization discrepancies are removed by correcting the shape of the star, and only small deviations due to the space-time quadrupole deformations are left, as noted  previously by \citet{PMN18}.

\begin{figure}
\resizebox{\hsize}{!}{\includegraphics{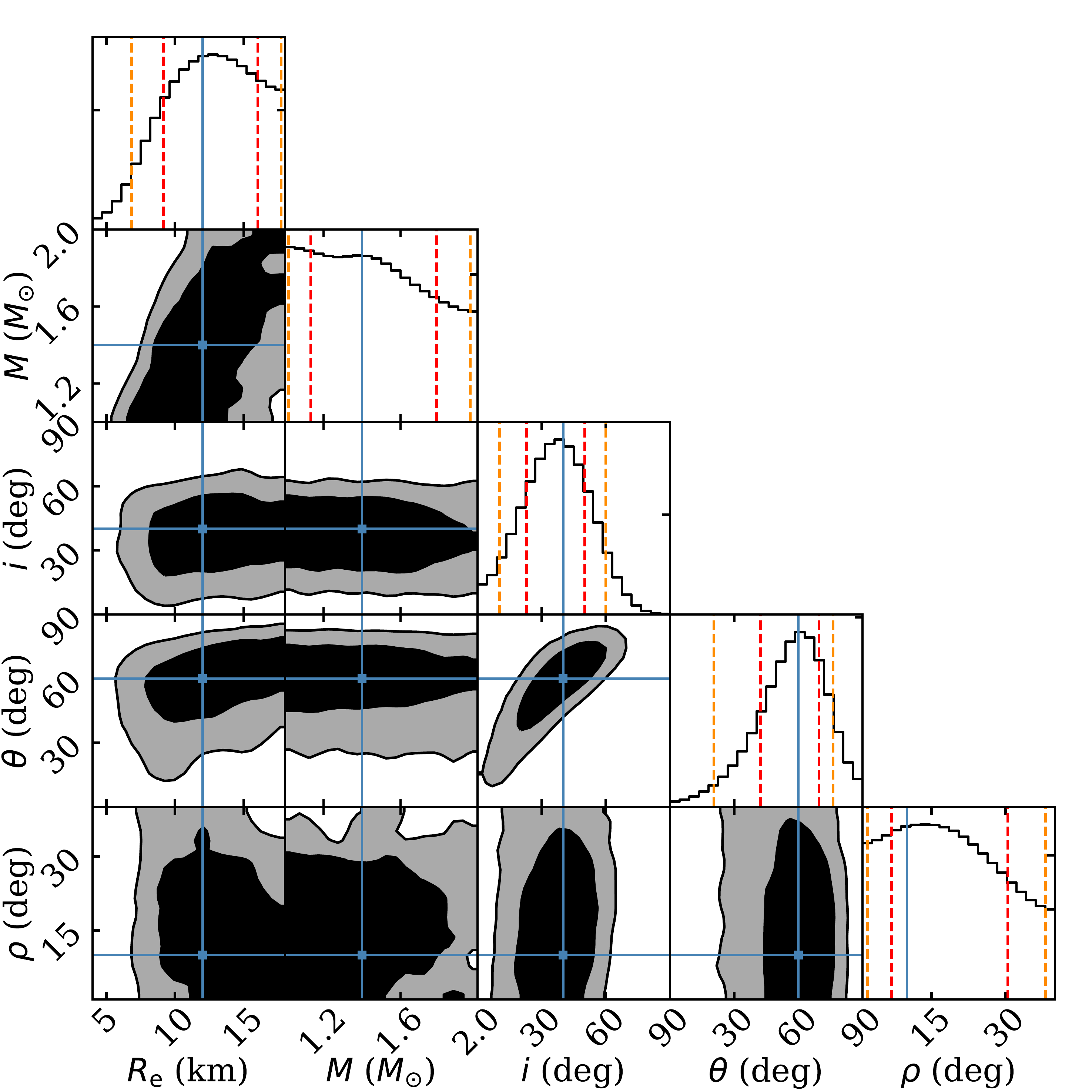}}
\caption{
Posterior distributions for an oblate fit to an oblate star assuming 15\degr\ standard deviation in the measured PA. 
In the two-dimensional posterior distributions the 68\% (in black) and the  95\% (in gray) highest posterior density credible regions are shown. 
In the one-dimensional posterior distributions the dashed red contour shows a 68\% and the dashed orange contour a 95\% highest posterior density credible interval. 
The blue lines show the input values. 
} 
\label{fig:oblobl_he}
\end{figure}

\subsection{Determining geometrical parameters from polarization data}\label{sec:res1a}

New constraints on the geometrical parameters (and thus also on the mass and radius) of the accreting millisecond pulsars are expected to be obtained with the upcoming polarimeters and X-ray missions like IXPE and eXTP. 
Therefore, we  made simple estimates to predict how accurately the main geometrical parameters, namely the observer inclination $i$ and the spot co-latitude $\theta$, can be measured.

Using our model for polarized radiation from two antipodal hotspots (presented in Sect. \ref{sec:theory}), we produced synthetic PA profile data, assuming a fixed uncertainty of the PA in all the phase bins. 
In addition, we considered profiles only at 5 keV energy because the PA profile is almost independent of energy.
The PA profiles were fitted using an affine invariant ensemble sampler with \textsc{emcee} \citep{emcee} package implemented in \textsc{python}. 
We kept the mass $M$, the equatorial radius $\req$, the observer inclination $i$, the spot co-latitude $\theta$, and the angular radius of the spot $\rho$ as free parameters, and assumed either a constant 15\degr\ or a 2\degr\ standard deviation in the measured PA.
The latter error is achievable, for instance, with \text{eXTP} in the energy range 2--10~keV, assuming brightness of 0.1 Crab, 4\% PD, 1~Ms exposure, and a negligible amount of background counts.
The 15\degr\ error would correspond to a 1~ks exposure but otherwise the same assumptions.
In the case of IXPE the errors are expected to be the same order of magnitude.

Instead of fitting PA profiles, it is preferable to use Stokes parameters because they are additive and their distribution is very well described by a bivariate normal distribution, even though they are neither the directly observed quantities from X-ray polarimeters \citep[see][]{KCB15}.
For simplicity, here we assumed that the errors in the measured PA would also be normally distributed, which is not true in general. 
In actual observations the error in PA will also be much higher at those pulsar phases when the polarization degree is closer or below the detection limit (which happens when the flux is at the maximum). 
A more accurate study of simulated observations, using Stokes parameters and the energy response of the IXPE detector, will be made in a forthcoming publication \citep{SLK20}. 

\begin{figure}
\resizebox{\hsize}{!}{\includegraphics{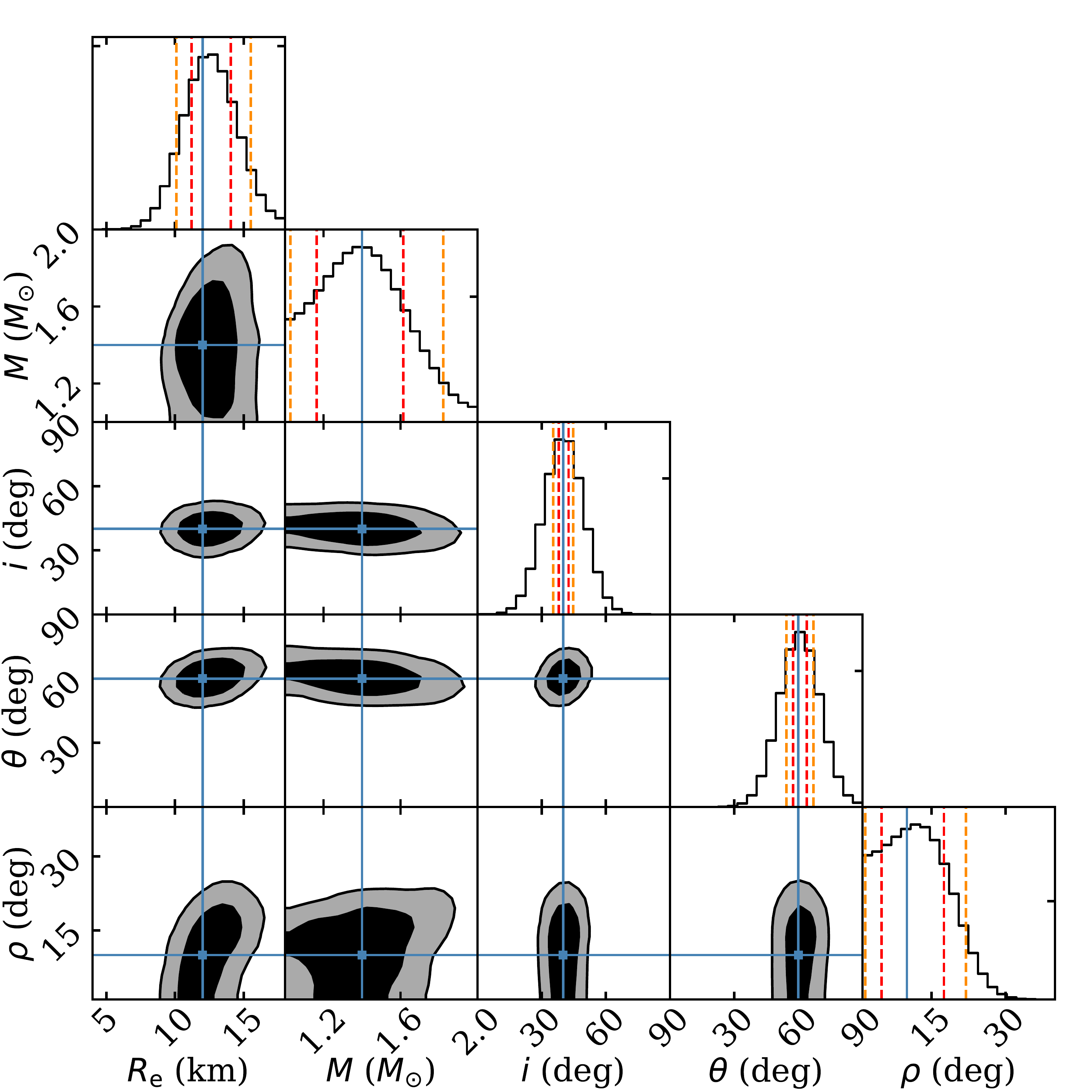}}
\caption{
Posterior distributions for an oblate fit to an oblate star assuming $2\degr$ standard deviation in the measured PA. 
The colors, contours, and other symbols are the same as in Fig.~\ref{fig:oblobl_he}.
} 
\label{fig:oblobl_le}
\end{figure}

The synthetic data were created assuming a rapidly rotating star ($\nu_* = 600$ Hz to emphasize the difference between an oblate and a spherical star) with $M = 1.4 ~\msun$, $\req = 12$ km, $i = 40 \degr$, $\theta = 60 \degr$, and $\rho = 10 \degr$.
When fitting the data, we also assumed a very conservative time resolution and used only ten phase bins. 
We treated the phase shift as a nuisance parameter. 
Instead of marginalized likelihoods over all phase shifts, we used a bisection method to maximize the likelihood, which is much faster, while the difference between these two methods is negligible \citep{SNP18}.

\begin{table*}
\begin{center}
\begin{minipage}{200mm}
\caption{Most probable values and 68\% and 95\% credible intervals  for three different simulations with synthetic data.}
\label{table:conflimits}
\begin{tabular}{l c c c c c}
\hline\hline
Quantity & 95\% lower limit & 68\% lower limit & Most probable value& 68\% upper limit & 95\% upper limit \\ \hline   
\multicolumn{6}{c}{Oblate fit to oblate star (large error in the measured PA)} \\ 
$\req$ (km) & $6.85$ & $9.16$ & $12.6$ & $16.0$ & $17.7$ \\
$M$ ($\msun$) & $1.02$ & $1.13$ & $1.44$ & $1.79$ & $1.96$ \\
$i$ ($\deg$) & $10$ & $23$ & $37$ & $50$ & $60$ \\
$\theta$ ($\deg$) & $21$ & $42$ & $59$ & $70$ & $76$ \\
$\rho$ ($\deg$) & $2.0$ & $6.9$ & $18$ & $30$ & $38$ \\
     \multicolumn{6}{c}{Oblate fit to oblate star (small error in the measured PA)} \\ 
$\req$ (km) & $10.1$ & $11.2$ & $12.5$ & $14.0$ & $15.5$ \\
$M$ ($\msun$) & $1.03$ & $1.17$ & $1.39$ & $1.61$ & $1.82$ \\
$i$ ($\deg$) & $35$ & $38$ & $40$ & $43$ & $45$ \\
$\theta$ ($\deg$) & $54$ & $57$ & $61$ & $64$ & $67$ \\
$\rho$ ($\deg$) & $1.6$ & $4.9$ & $12$ & $17$ & $22$ \\
      \multicolumn{6}{c}{Spherical fit to oblate star (small error in the measured PA)} \\ 
$\req$ (km) & $10.3$ & $11.6$ & $13.5$ & $16.0$ & $17.6$ \\
$M$ ($\msun$) & $1.04$ & $1.20$ & $1.44$ & $1.70$ & $1.92$ \\
$i$ ($\deg$) & $36$ & $38$ & $40$ & $42$ & $44$ \\
$\theta$ ($\deg$) & $50$ & $52$ & $55$ & $57$ & $59$ \\
$\rho$ ($\deg$) & $2.0$ & $6.2$ & $16$ & $23$ & $28$ \\
 \hline
  \end{tabular}
 \end{minipage}
\end{center}
\tablefoot{The correct values of the parameters are $\req = 12$~km, $M = 1.4 ~\msun$, $i = 40 \degr$, $\theta = 60 \degr$, and $\rho = 10 \degr$.
}  
\end{table*}

\begin{figure}
\resizebox{0.9\hsize}{!}{\includegraphics{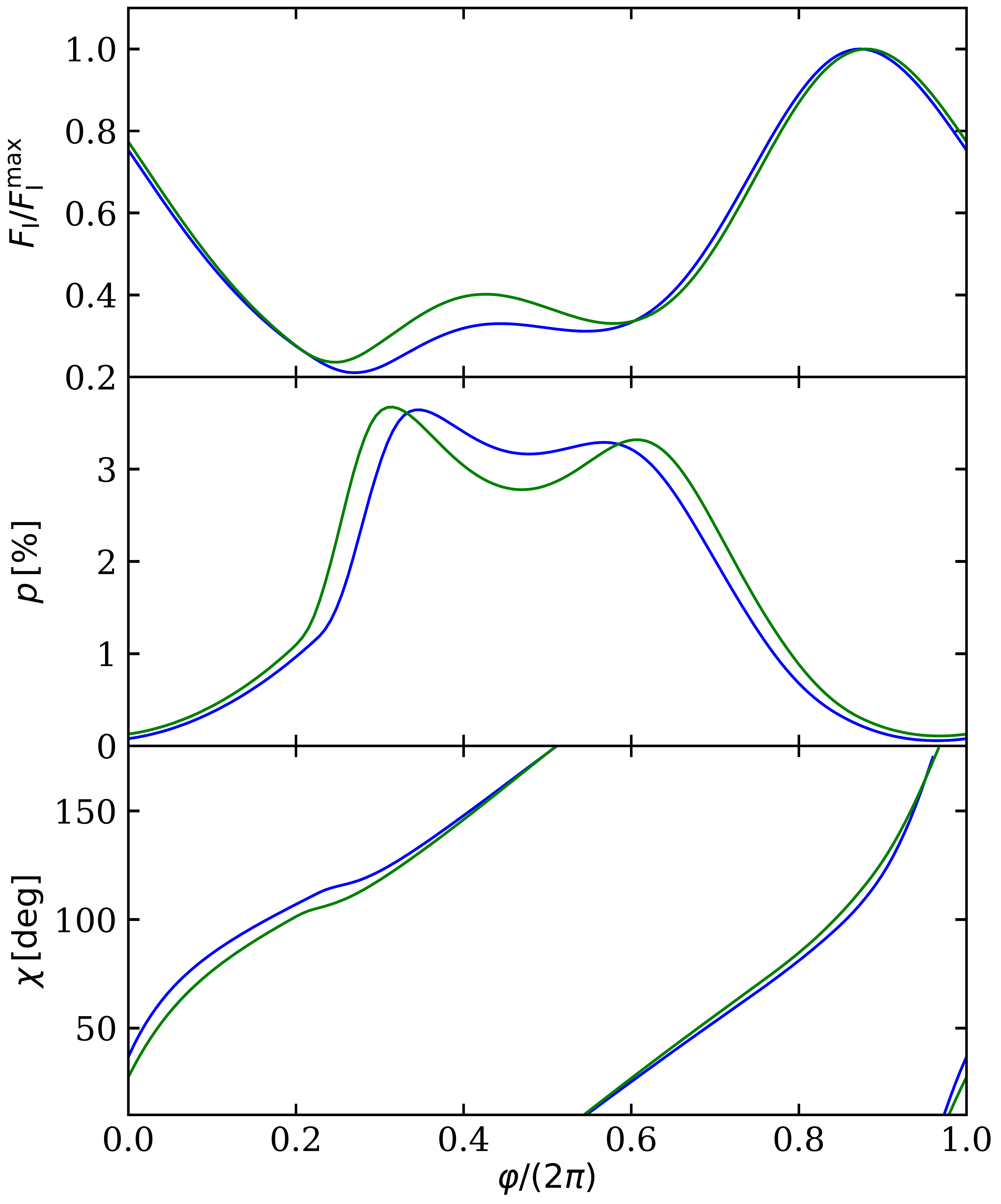}} 
    \caption{Flux (\textit{top panel}), PD (\textit{middle panel}), and PA (\textit{bottom panel}) as a function of observed  phase for an oblate (in blue) and spherical (in green) NS with identical model parameters at 5~keV (given in Sect.~\ref{sec:res1a}). 
    For the spherical star corresponding non-rotating mass and radius are used: $M_{0}=1.394 ~\msun$ and $R_{0} = 11.43$ km.}
    \label{fig:oblsph_comp}
\end{figure}

The results of these fits are shown in Figs.~\ref{fig:oblobl_he} and \ref{fig:oblobl_le}. 
The credible intervals are also presented in Table~\ref{table:conflimits}.
As expected, we found a significant improvement in the parameter constraints assuming a smaller error in PA (i.e., longer exposure time of the detector). 
The 68\% credible interval for the inclination angle $i$ changes from $23\degr$--$~50\degr$ to $38\degr$--$~43\degr$ and for spot co-latitude from $42\degr$--$~70\degr$ to $57\degr$--$~64\degr$. 
In the case of short exposure, the mass and radius are not very well constrained.
However, in the case of a long exposure we also get limits  for radius $\req = 12.5^{+1.5\,(3.0)}_{-1.3\,(2.4)}$~km  and even for mass $M = 1.39^{+0.22\,(0.43)}_{-0.22\,(0.36)} ~\msun$ 
with 68\%  (95\%) limits. 
This is mainly due to the influence of mass and radius on the moment when the second spot becomes visible and starts to contribute to the PA profile. 
The size of the spot $\rho$ is also not very well constrained, although with a long exposure we get an upper limit of 22\degr\ with 95\% certainty.
These limits should be taken as estimates because the model describes the synthetic data perfectly, which is not the case with real observations.
We also note that much tighter constraints for the mass and radius are expected when combining polarization measurements with the modeling of the pulse profiles, i.e., flux (which is not considered here).
The most essential benefit of polarization data is expected to be in breaking the degeneracy in $i$ and $\theta$ \citep{VP04}.

\begin{figure}
\resizebox{\hsize}{!}{\includegraphics{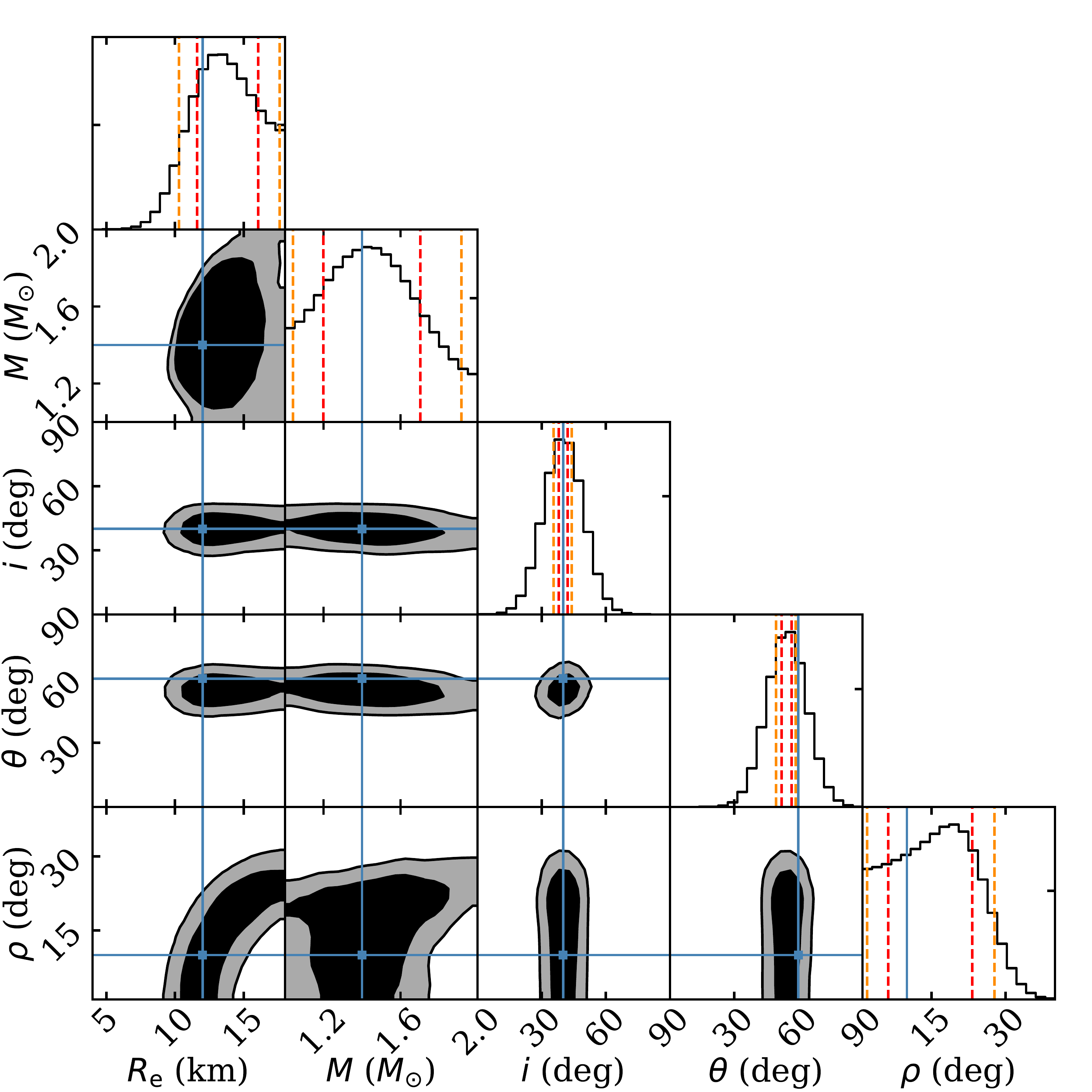}}
\caption{
Posterior distributions for a spherical fit to an oblate star assuming $2\degr$ standard deviation in the measured PA. 
The colors, contours, and other symbols are the same as in Fig. \ref{fig:oblobl_he}.
} 
\label{fig:oblsph_le}
\end{figure}

\subsection{Effects of oblateness on the polarization}\label{sec:res1b}

To test the effect of oblateness on the PA, we also fitted the synthetic PA profile (for oblate star) using an S+D model for a spherical star. 
The free parameters of our model were the same as in  Sect. \ref{sec:res1a}.
The synthetic PA profile is shown in Fig.~\ref{fig:oblsph_comp} with a blue curve, and the corresponding profile with a spherical star is shown with a green curve (the former being the same profile    used in Sect.~\ref{sec:res1a}).
As in Fig.~\ref{fig:arc_agm_obl_sph}, we see again evident differences between the two models.  

The results of the fitting are shown in Fig.~\ref{fig:oblsph_le}  (spherical fit to an oblate star) for a $1\sigma$ error of $2\degr$ in the measured PA, and they can be compared with those obtained using the correct model with the same exposure time in Fig.~\ref{fig:oblobl_le}.
The credible intervals are  also found from Table~\ref{table:conflimits}.
The results show no bias when the correct model is used, but with an incorrect model for the star shape we see an obvious bias in the spot co-latitude $\theta$  giving too small values for the angle.
The correct value is outside of the measured 95\% credible interval.
This is likely explained by the co-latitude appearing in terms of $\lambda = \theta-\eta$ in PA Eq. \eqref{eq:chiprime}. 
Fitting a spherical model to an oblate star tends to find a correct $\lambda$ rather than correct $\theta$.
It shows that the effects of oblateness are important to include when calculating the PA profiles, at least for stars rotating as fast as $\nu_* = 600$~Hz. 

In addition, we see a small but noticeable shift in the measured radius $\req$ towards higher values (even though $\req$ should be smaller for a corresponding non-rotating star as seen in Sect. \ref{sec:geometry}). 
However, the correct value is still inside the 68\% credible interval. 
A minor dependence of the PA profile on $\req$ is not surprising because the radius affects the time interval when the second spot is visible. 
The visibility condition is also different for oblate and spherical stars, and thus it can be responsible for biasing the constraints in the radius.

\section{Conclusions}\label{sec:conclusions}

We  developed the formalism to compute the light curves of Stokes parameters observed from a rapidly rotating NS in the OS approximation. 
We derived analytical formulae describing the rotation of the polarization plane as a result of special and general relativistic effects accounting for the oblate shape of a NS. 
The expression for the PA is summarized in Eqs.\,\eqref{eq:pa_rvm}, \eqref{eq:chi1}, \eqref{eq:chiprime}, and \eqref{eq:chi_tot}.
In the case of a spherical star, our results simplify to those presented in \citet{VP04} and \citet{poutanen20b}.
We also verified the accuracy and correctness of our formalism by comparing the resulting polarization pulse profiles to numerical calculations computed with the general relativistic polarized ray-tracing code \textsc{arcmancer}.

In addition, we estimated that the accuracy of a few degrees in the measured observer inclination and spot co-latitude can be obtained from the upcoming X-ray polarization measurements, assuming a correct theoretical model and using data with high precision.
We caution that the resulting system parameters can be seriously biased if a spherical star shape is erroneously used to compute the profiles emitted by an oblate NS.

We also note that we only considered the simple case with two antipodal hotspots assuming a dipolar magnetic field configuration. 
The presented model, however, can also be applied to more complicated spot configurations, such as recently inferred in MSPs based on the observations of PSR~J0030+0451 by the Neutron star Interior Composition ExploreR \citep{BWH_nicer19,MLD_nicer19,RWB_nicer19}.
In that case, the polarized pulse profiles would show a more complicated structure.
Because the non-antipodal spots have largely different PAs, the observed total PA would show abrupt breaks when any of the spots eclipses. 
Future polarimetric observations will thus be able to test the magnetic field geometry in MSPs.


\begin{acknowledgements}
This research was supported by the Magnus Ehrnrooth Foundation, the University of Turku Graduate School in Physical and Chemical Sciences (TS), the Academy of Finland grants 322779 (JP) and 333112 (VL, JP), and the Russian Science Foundation grant 20-12-00364 (VL, JP). 
The computer resources of the Finnish IT Center for Science (CSC) and the FGCI project (Finland) are acknowledged. 
\end{acknowledgements}

\bibliographystyle{aa}
\bibliography{ns_pol}

\appendix

\section{Alternative derivation for the general relativistic correction of PA}
\label{sec:appendix}
 
The general relativistic correction of PA for the Schwarzschild metric can also be derived using the common vector formalism presented in \citet{poutanen20b}. 
Here we reproduce the necessary equations from their Sect. 3.2.2, but considering an oblate NS instead of a spherical one.
The common vector perpendicular to both photon momentum and the spot velocity vector in both frames is  
\beq \label{eq:comm_vector_rel}
\unit{M} & =&  \frac{\unit{k}_0 \times \unit{\beta}  }{|\unit{k}_0 \times \unit{\beta} |}  
= \frac{\unit{k}^{\prime}_0 \times \unit{\beta}  }{|\unit{k}^{\prime}_0\times \unit{\beta} |}  \\
& =&\frac{1}{\sin\xi \sin\psi} \left( \begin{array}{c} 
-\cos \phi\ [\cos i \sin \alpha + \cos\theta \sin(\psi-\alpha)] \\
-\sin \phi\ [\cos i \sin \alpha + \cos\theta \sin(\psi-\alpha)] \\
 \sin i \cos \phi  \sin \alpha + \sin\theta \sin(\psi-\alpha)  
\end{array}  \right) .    \nonumber
\eeq
The unit vector of the photon momentum in the spot comoving frame comes from the Lorentz transformation: 
\beq 
\unit{k}^{\prime}_0&=& \delta \left[ \unit{k}_0 - \gamma\beta \unit{\beta} + \gamma-1) \unit{\beta} (\unit{\beta}\cdot \unit{k}_0 ) \right] .
\eeq 
These expressions do not depend on the NS shape.
The polarization basis related to photon momentum in the comoving frame $\unit{k}_0'$ and the surface normal is (for an oblate star) 
\be\label{eq:pout47}
\unit{e}_1'^{0} = \frac{\unit{n}-\cos{\sigma'}\  \unit{k}_0'}{\sin{\sigma'}},\qquad 
\unit{e}_2'^{0} = \frac{\unit{k}_0' \times \unit{n}}{\sin{\sigma'}} .
\ee
The angle between vector $\unit{M}$ and the basis vector $\unit{e}_1'^{0}$ is 
\beq\label{eq:pout48}
\cos{\chi'_{\mathrm{M}}}&=&\unit{M} \cdot \unit{e}_1'^{0} = \frac{C}{\sin\xi\sqrt{1-\delta^{2}\cos^{2}}\sigma} , \\
\label{eq:pout49}
\sin{\chi'_{\mathrm{M}}}&= & - \unit{M} \cdot \unit{e}_2'^{0} 
= \frac{\cos \sigma}{\sin \xi \sqrt{1-\delta^{2}\cos^{2}\sigma}}\frac{\cos \xi - \beta}{1 - \beta \cos \xi},
\eeq
where
\be
C=-\frac{\sin \alpha}{\sin\psi} \cos\zeta  + \frac{\sin (\psi-\alpha)}{\sin\psi} \sin \eta .
\ee
Therefore, we get
\be\label{eq:pout50}
\tan{\chi'_{\mathrm{M}}} = \frac{\cos \sigma}{C} \frac{\cos \xi - \beta}{1 - \beta \cos \xi} .
\ee

The polarization basis in the laboratory frame
is
\be\label{eq:pout51}
\unit{e}_1^{0} = \frac{\unit{n}-\cos{\sigma}\  \unit{k}_0}{\sin{\sigma}},\qquad 
\unit{e}_2^{0} = \frac{\unit{k}_0 \times \unit{n}}{\sin{\sigma}} .
\ee 
Now we can evaluate the angle $\chi_{\mathrm{M}}$ between the common vector $\unit{M}$ and the projection of the stellar normal on the sky as viewed in that frame: 
\beq\label{eq:pout52}
\cos{\chi_{\mathrm{M}}}& = & \unit{e}_1^{0} \cdot \unit{M} = 
\frac{C}{\sin \sigma \sin \xi} ,  \\
\label{eq:pout53}
\sin{\chi_{\mathrm{M}}} & = &  \unit{e}_2^{0} \cdot \unit{M} =  
-\frac{\cos \sigma \cos\xi }{\sin \sigma \sin \xi} ,
\eeq
and 
\be\label{eq:pout54}
\tan{\chi_{\mathrm{M}}} = 
- \frac{\cos \sigma \cos\xi }{C} .
\ee
Thus, the total general relativistic correction angle $\chi'=\chi'_{\mathrm{M}}+\chi_{\mathrm{M}}$ can be found as 
\beq\label{eq:pout55}
\tan\chi' & = &  \frac{\tan{\chi'_{\mathrm{M}}}+\tan{\chi_{\mathrm{M}}}}{1-\tan{\chi'_{\mathrm{M}}}\tan{\chi_{\mathrm{M}}}}  \\
&=& - \frac{\beta \cos \sigma\ C \sin^2\xi}{C^2+ \cos^2\sigma \cos^2\xi - \beta\cos\xi (C^2+\cos^2\sigma)}. \nonumber 
\eeq
It is easy to show that $C^2+\cos^2\sigma=\sin^2\xi$, so we finally get 
\be\label{eq:pout55b}
\tan\chi' = - \frac{\beta \cos \sigma\ C}{  \sin^2\sigma  - \beta\cos\xi } ,
\ee
which is equivalent to Eq.\,\eqref{eq:chiprime}.

\end{document}